\documentclass[journal]{IEEEtran}

\usepackage{cite,bm,tikz,amsfonts,amssymb,graphicx,pgfplots,multirow,epstopdf,algorithm,algorithmicx,algpseudocode}
\usepackage[acronym]{glossaries}
\usetikzlibrary{arrows.meta,decorations.pathreplacing,plotmarks,shapes,arrows,positioning}
\pgfplotsset{compat=1.12}
\pgfplotsset{minor grid style={dotted,white!70!black}}
\pgfplotsset{every tick label/.append style={font=\scriptsize}}
\let\originalleft\left
\let\originalright\right
\renewcommand{\left}{\mathopen{}\mathclose\bgroup\originalleft}
\renewcommand{\right}{\aftergroup\egroup\originalright}

\newcommand{\Qone}{\mathbf Q_1}
\newcommand{\Qtwo}{\mathbf Q_2}
\newcommand{\define}{\triangleq}
\newcommand{\modset}{\mathcal X}
\newcommand{\LPN}[1]{\theta_{#1}^\mathrm{L}}
\newcommand{\coreDrift}[2]{\theta_{#1,#2}^\mathrm{C}}
\newcommand{\polDrift}[2]{\theta_{#1,#2}^\mathrm{P}}
\newcommand{\varLPN}{\sigma_\mathrm{L}^2}
\newcommand{\varCoreDrift}{\sigma_\mathrm{C}^2}
\newcommand{\varPolDrift}{\sigma_\mathrm{P}^2}
\newcommand{\varN}[1]{\sigma_{#1}^2}
\newcommand{\varNtilde}[1]{\bar\sigma_{#1,k}^2}
\newcommand{\snr}{\mathrm{SNR}_\mathrm{b}}
\newcommand{\snrinline}{\mathrm{SNR}_\mathrm{b}}
\newcommand{\Es}{E_\mathrm{s}}
\newcommand{\OHpilot}{\mathrm{OH}_\mathrm{P}}
\newcommand{\OHFEC}{\mathrm{OH}_\mathrm{FEC}}

\DeclareMathOperator*{\argmax}{argmax}
\DeclareMathOperator*{\diag}{diag}

\newacronym{cpe}{CPE}{carrier-phase estimation}
\newacronym{mcf}{MCF}{multicore fiber}
\newacronym{sdm}{SDM}{space-division multiplexed}
\newacronym{dsp}{DSP}{digital signal processing}
\newacronym{qam}{QAM}{quadrature amplitude modulation}
\newacronym{dp}{DP}{dual-polarization}
\newacronym{dpmqam}{DP-$M$QAM}{dual-polarization $M$-ary quadrature amplitude modulation}
\newacronym{dpqpsk}{DP-QPSK}{\gls{dp} quadrature phase-shift keying}
\newacronym{snr}{SNR}{signal-to-noise ratio}
\newacronym{lpn}{LPN}{laser phase noise}
\newacronym{bps}{BPS}{blind phase search}
\newacronym{ber}{BER}{bit error rate}
\newacronym{oh}{OH}{overhead}
\newacronym{awgn}{AWGN}{additive white Gaussian noise}
\newacronym{eks}{EKS}{extended Kalman smoothing}
\newacronym{fec}{FEC}{forward error correction}
\newacronym{pdf}{PDF}{probability density function}
\newacronym{pccpe}{PC-CPE}{per-channel \gls{cpe}}
\newacronym{mscpe}{MS-CPE}{master--slave \gls{cpe}}
\newacronym{jccpe}{JC-CPE}{joint-channel \gls{cpe}}
\newacronym{qpsk}{QPSK}{quadrature phase-shift keying}
\newacronym{cfo}{CFO}{carrier-frequency offset}
\newacronym{lo}{LO}{local oscillator}
\newacronym{ecl}{ECL}{external cavity laser}
\newacronym{wdm}{WDM}{wavelength-division multiplexing}
\newacronym{fgk}{FGK}{factor-graph based EKS}
\newacronym{map}{MAP}{maximum \textit{a posteriori}}
\newacronym{ser}{SER}{symbol error rate}
\newacronym{gmi}{GMI}{generalized mutual information}
\newacronym{lms}{LMS}{least mean square}
\newacronym{rde}{RDE}{radially-directed equalizer}
\newacronym{air}{AIR}{achievable information rate}

\begin{document}

\title{Pilot-Aided~Joint-Channel~Carrier-Phase\\Estimation~in~Space-Division~Multiplexed\\Multicore~Fiber~Transmission}

\author{Arni~F.~Alfredsson,~\IEEEmembership{Student~Member,~IEEE},~Erik~Agrell,~\IEEEmembership{Fellow,~IEEE},\\~Henk~Wymeersch,~\IEEEmembership{Member,~IEEE},~Benjamin~J.~Puttnam,~\IEEEmembership{Member,~IEEE},~Georg~Rademacher,~\IEEEmembership{Member,~IEEE},\\Ruben~S.~Lu\'is,~and~Magnus~Karlsson,~\IEEEmembership{Senior~Member,~IEEE,~Fellow,~OSA}%
	\thanks{A. F. Alfredsson, E. Agrell, and H. Wymeersch are with the Department of Electrical Engineering, Chalmers University of Technology, SE-41296 G\"{o}teborg, Sweden (e-mail: arnia@chalmers.se; agrell@chalmers.se; henkw@chalmers.se).
		
	M. Karlsson is with the Photonics Laboratory, Department of Microtechnology and Nanoscience, Chalmers University of Technology SE-41296 G\"{o}teborg, Sweden (e-mail: magnus.karlsson@chalmers.se).
	
	B. J. Puttnam, G. Rademacher, and R. S. Lu\'{i}s are with the Photonic Network System Laboratory, National Institute of Information and Communications Technology, 4-2-1 Nukui-Kitamachi, Koganei, Tokyo 184-8795, Japan (e-mail: ben@nict.go.jp; georg.rademacher@nict.go.jp; rluis@nict.go.jp).
	
	This work was supported by the Swedish Research Council (VR), Grants 2013-5642 and 2014-6138.
	}
}
\maketitle

\begin{abstract}
	\boldmath
	The performance of pilot-aided joint-channel \gls{cpe} in space-division multiplexed \gls{mcf} transmission with correlated phase noise is studied. To that end, a system model describing uncoded \gls{mcf} transmission where the phase noise comprises a common laser phase noise, in addition to core- and polarization-specific phase drifts, is introduced. It is then shown that the system model can be regarded as a special case of a multidimensional random-walk phase-noise model. A pilot-aided \gls{cpe} algorithm developed for this model is used to evaluate two strategies, namely joint-channel and per-channel \gls{cpe}. To quantify the performance differences between the two strategies, their respective phase-noise tolerances are assessed through Monte Carlo simulations of uncoded transmission for different modulation formats, pilot overheads, laser linewidths, numbers of spatial channels, and degrees of phase-noise correlation across the channels. For 20 GBd transmission with 200 kHz combined laser linewidth and 1\% pilot overhead, joint-channel \gls{cpe} yields up to 3.4 dB improvement in power efficiency or 25.5\% increased information rate. Moreover, through \gls{mcf} transmission experiments, the system model is validated and the strategies are compared in terms of bit-error-rate performance versus transmission distance for uncoded transmission of different modulation formats. Up to 21\% increase in transmission reach is observed for 1\% pilot overhead through the use of joint-channel \gls{cpe}.
\end{abstract}

\glsresetall

\section{Introduction}
\label{sec:intro}
In response to the ever-increasing throughput demands on fiber-optical networks, \gls{sdm} systems have become a topic of interest worldwide \cite{sdm:richardson13}. They are believed to have the potential to meet the demands in a cost-effective manner through, e.g., the integration of optical hardware components, sharing of \gls{dsp} resources \cite{6317137}, specialized detection techniques \cite{7121584}, and the use of spatial superchannels \cite{6533237}. Moreover, in order to maximize the information rate of the system, multilevel modulation formats, such as
\gls{dpmqam}\glsunset{dp}\glsunset{qam}
or more advanced multidimensional formats \cite{Karlsson:17}, are being increasingly utilized. However, such higher-order formats typically come at the cost of higher \gls{snr} requirements. They can also increase sensitivity to various transmission impairments, in particular \gls{lpn}, which calls for effective \gls{cpe}. 
In addition, symbol-rate optimization implemented through subcarrier multiplexing has been a topic of interest in recent years as gains in transmission reach have been observed by using symbol rates on the order of 2--6 GBd, which is much lower than standard symbol rates \cite{7376999}. The impact of \gls{lpn} increases as the symbol rate is decreased, and thus, powerful \gls{cpe} is even more crucial for symbol-rate-optimized systems.

Traditionally, \gls{cpe} in optical transmission systems is performed on a per-channel basis, using blind methods such as the Viterbi--Viterbi algorithm \cite{1056713} or \gls{bps} \cite{4814758}. However, due to the $\pi/2$ rotation invariance that is inherent to the most commonly used modulation formats, blind methods suffer from ambiguity in the estimated carrier phase and are thus susceptible to cycle slips, which can lead to bursts of errors. Differential encoding can be used to convert the burst errors into a finite number of errors, but it increases the \gls{ber} by a constant factor compared to Gray coding in the absence of cycle slips \cite{4298982}. Alternatively, the \gls{cpe} can be carried out with the help of pilot symbols whose phases are known unambiguously. This greatly reduces the probability of cycle slips \cite{Cheng:13} and eliminates the need for differential encoding. Although this comes at the cost of reduced information rate, pilot-aided \gls{cpe} methods have garnered attention in recent years due to their high performance \cite{6533070,7341655,Zhu:17,Le:14}.

\Gls{cpe} has been experimentally demonstrated for various superchannel transmission scenarios, e.g., \gls{sdm} transmission using \glspl{mcf} \cite{6317137} or multimode fibers \cite{6517220}, and wavelength-division multiplexed transmission using frequency combs \cite{LarsECOC:17} or electrically generated subcarriers \cite{6290331}. These demonstrations were possible due to the spatial correlation in the phase noise that is inherent to these systems \cite{6317137,6517220,LarsECOC:17,6290331}. In particular, for \gls{sdm} transmission using \glspl{mcf}, the \gls{lpn} will be common among the spatial channels if all cores share a light source on the transmitter side and a \gls{lo} on the receiver side \cite{6317137}. However, temperature variations, other external perturbations, and imperfections or certain properties in system components will cause phase drifts that are specific to cores and polarizations \cite{7183869,6317137}. These phase drifts are normally orders of magnitude slower than the \gls{lpn}, and thus, the phase noise will have a high degree of spatial correlation across the channels.

The correlation can be exploited to lower the required computational complexity in \gls{dsp} through optical techniques \cite{7121584,mazur_jlt18}, or with the help of \gls{dsp}-based methods, such as \gls{mscpe} \cite{6317137}.
These strategies rely on \gls{cpe} using a single spatial channel, whose outcome is shared amongst all spatial channels. Although this may substantially reduce the required \gls{cpe} resources, any phase differences between the channels will reduce their effectiveness. In contrast, by performing \gls{jccpe} where all channels are used collectively, phase-noise tolerance can be improved. This can be used to benefit system performance in terms of power efficiency, information rate, hardware requirements, or transmission reach, at the cost of added computational complexity. A comparison between pilot-aided \gls{jccpe}, \gls{pccpe}, and \gls{mscpe} was made in \cite{agrell_cleo18,aplp_puttnam18} in terms of \gls{ber} versus transmission distance for \gls{mcf} transmission of \gls{dp}-16\gls{qam}, and it was shown that \gls{jccpe} can cope more effectively than \gls{mscpe} with phase differences between the cores.

Algorithms that perform \gls{jccpe} have been extensively investigated for wireless multiple-input multiple-output transmission \cite{1400262,7080901}. Furthermore, we recently proposed pilot-aided algorithms that perform \gls{jccpe} for optical transmission using \gls{fec} in the presence of arbitrarily correlated phase noise for any number of channels \cite{alfredsson:tcom18_arxiv}. We further showed that they can significantly outperform the typical \gls{cpe} approach of using \gls{bps} on a per-channel basis in terms of post-\gls{fec} \gls{ber} performance.

In this paper, we investigate the performance of \gls{jccpe} for \gls{sdm} transmission through uncoupled-core \glspl{mcf}. The contributions are summarized as follows: (i) We introduce a general phase-noise model for uncoupled-core \gls{mcf} transmission that comprises a common \gls{lpn}, in addition to core- and polarization-specific phase drifts that are independent of each other and decorrelate the common phase noise among the spatial channels. (ii) We show that this phase-noise model can be regarded as a multidimensional random walk and utilize a pilot-aided algorithm to perform \gls{jccpe} for arbitrarily correlated phase noise and any number of channels. (iii) Using Monte Carlo simulations, we compare the performance differences between two strategies, namely \gls{jccpe} and \gls{pccpe}, in terms of the resulting power efficiency, information rate, and laser-linewidth requirements of the system. This part of the paper refines preliminary results that were presented in \cite{alfredsson2017ofc}. (iv) Finally, we experimentally validate the system model and evaluate the two strategies in terms of transmission reach for \gls{dpqpsk}, \gls{dp}-16\gls{qam}, and \gls{dp}-64\gls{qam} transmission using different pilot \glspl{oh}. This part extends the \gls{pccpe} and \gls{jccpe} comparison presented in \cite{alfredsson2018sum}.

\section{System Model}
\label{sec:sysmodel}
Consider uncoded single-wavelength \gls{dp} transmission in $D/2$ cores resulting in a total of $D$ spatial channels where, without loss of generality, $D$ is assumed to be an even integer. The transmitted symbol block in each channel is modelled as a vector of $N$ independent random variables, where every random variable corresponding to a data symbol is drawn uniformly from a set $\modset$ of constellation points that corresponds to the used modulation format. Moreover, the constellation is normalized such that the mean of the constellation points is zero and the average symbol energy is $E_s$.

All signal distortions are assumed to have been ideally compensated with the exception of phase noise and amplified spontaneous emission noise, which are approximated as a random walk and \gls{awgn}, resp. Moreover, the adaptive equalization is assumed to have been carried out using a phase-immune equalizer, such as a \gls{rde} \cite{4909145}. As already mentioned, the phase noise can be highly correlated, albeit not identical, across the cores and polarizations in an \gls{mcf} system where all cores share the light-source and \gls{lo} lasers \cite{7183869,6317137}. In the absence of a model that accurately describes the phase-noise statistics across all channels, we resort to a simplified model in which the phase noise comprises multiple components. The dominant component is common to all the spatial channels and is assumed to contain the combined \gls{lpn} of the light source and \gls{lo}. The other components, which account for effects that decorrelate the phase noise across the channels, are core- and polarization-specific phase drifts that are statistically independent of each other. Also assuming one sample per symbol, the discrete-time baseband model is written as
\begin{equation}
	r_{i,k} = s_{i,k}e^{j\theta_{i,k}}+n_{i,k},\label{eq:sysmodel}
\end{equation}
where $k=1,\dots,N$ is a time index and $i=1,\dots,D$ is a channel index. The received signals, transmitted symbols, and \gls{awgn} samples are denoted with $r_{i,k}$, $s_{i,k}$, and $n_{i,k}$, resp. Each $n_{i,k}$ is the realization of a zero-mean Gaussian random variable with variance $\sigma_i^2$ per real dimension, which can be different for each channel. A specific mapping between the channel indices and polarization--core combinations is assumed. Denoting polarization $w\in\{\mathsf x,\mathsf y\}$ on the $j$th core with $(w,j)$, the channel indices $1,2,3,\dots,D-1,D$ correspond to $(\mathsf x,1),(\mathsf y,1),(\mathsf x,2),\dots,(\mathsf x,D/2),(\mathsf y,D/2)$. The sets of pilot and data symbol indices are denoted with $\mathcal P$ and $\mathcal D$, resp. Certain symbols within the transmitted blocks are designated as pilots such that if $(i,k)\in\mathcal P$ then $s_{i,k}=\rho_{i,k}$, where $\rho_{i,k}\in\mathbb C$ is known to the transmitter and receiver.

The phase noise $\theta_{i,k}$ is a sum of three statistically independent components, i.e., $\theta_{i,k}=\LPN{k}+\coreDrift{i}{k}+\polDrift{i}{k}$, where $\LPN{k}$, $\coreDrift{i}{k}$, and $\polDrift{i}{k}$ are the \gls{lpn}, core-specific phase drift, and polarization-specific phase drift, resp. The statistical nature of $\coreDrift{i}{k}$ and $\polDrift{i}{k}$ will be highly dependent on the system involved. For simplicity, all the components are approximated as random walks, i.e.,
\begin{align}
	\LPN{k}&=\LPN{k-1}+\Delta\LPN{k},\\
	\coreDrift{i}{k}&=\begin{cases}
		\coreDrift{i}{k-1}+\Delta\coreDrift{i}{k},&i\textnormal{ odd},\\
		\coreDrift{i-1}{k},&i\textnormal{ even},
	\end{cases}\\
	\polDrift{i}{k}&=\polDrift{i}{k-1}+\Delta\polDrift{i}{k},
\end{align}
for $k=2,\dots,N$, where $\LPN{1}$, $\coreDrift{i}{1}$, and $\polDrift{i}{1}$ are uniformly distributed on $[0,2\pi)$, and $\Delta\LPN{k}$, $\Delta\coreDrift{i}{k}$, and $\Delta\polDrift{i}{k}$ are zero-mean Gaussian random variables with variances $\varLPN$, $\varCoreDrift$, and $\varPolDrift$, resp. In particular, $\varLPN\define2\pi\Delta\nu T_\mathrm{s}$, where $\Delta\nu$ is the combined linewidth of the light-source and \gls{lo} lasers, and $T_\mathrm{s}$ is the symbol duration. Moreover, $\varCoreDrift$ and $\varPolDrift$ are defined in relation to $\varLPN$. These variances determine the speed of their corresponding phase drifts. The phase noise is statistically independent of the transmitted symbols and \gls{awgn}, and unknown to both the transmitter and receiver. Finally, the variances $\varLPN$, $\varCoreDrift$, $\varPolDrift$, and $(\varN{1},\dots,\varN{D})$ are assumed to be known to the receiver.

The phase noise can alternatively be described as a multidimensional random walk, i.e.,
\begin{equation}
	\bm\theta_k=\bm\theta_{k-1}+\Delta{\bm\theta}_k,\label{eq:random_walk}
\end{equation}
where $\bm\theta_1$ is uniformly distributed on $[0,2\pi)^{D}$ and $\Delta{\bm\theta}_k$ is a multivariate zero-mean Gaussian random variable with covariance matrix
\begin{equation}
	\mathbf Q=\begin{bmatrix}
		\Qone & \Qtwo & \Qtwo & \cdots & \Qtwo \\
		\raisebox{2pt}{$\Qtwo$} & \raisebox{2pt}{$\Qone$} & & & \vdots \\[-2pt]
		\raisebox{2pt}{$\Qtwo$} & & \ddots & & \vdots \\[-2pt]
		\vdots & & & \ddots & \raisebox{2pt}{$\Qtwo$} \\[2.7pt]
		\Qtwo & \cdots & \cdots & \Qtwo & \Qone
	\end{bmatrix}\in\mathbb R^{D\times D},\label{eq:rwcov}
\end{equation}
where
\begin{equation}
	\Qone = \begin{bmatrix}
		\varLPN+\varCoreDrift+\varPolDrift & \varLPN+\varCoreDrift \\[0.3em]
		\varLPN+\varCoreDrift & \varLPN+\varCoreDrift+\varPolDrift
	\end{bmatrix},~
	\Qtwo = \begin{bmatrix}
		\varLPN & \varLPN \\[0.3em]
		\varLPN & \varLPN
	\end{bmatrix}.
\end{equation}
The covariance matrix in \eqref{eq:rwcov} is specific to the structure of the phase noise in \eqref{eq:sysmodel}. The multidimensional random-walk description in \eqref{eq:random_walk} and \eqref{eq:rwcov} is utilized by the \gls{cpe} algorithm, which is detailed in the next section.

\section{\gls{cpe} Algorithm}
\label{sec:alg}

\Gls{jccpe} involves using all channels collectively in order to estimate the phase noise simultaneously across the channel domain. This strategy can be implemented using either blind or pilot-aided algorithms. If the phase noise is assumed identical in all channels, existing \gls{pccpe} algorithms be extended in a simple manner to perform estimate averaging across the channel domain, which will reduce the impact of the additive noise that corrupts the \gls{cpe}. However, to track interchannel phase drifts, more involved algorithms are required. Here, we make use of an iterative pilot-aided algorithm that is developed using the sum--product algorithm, which operates in a factor graph.
It approximates the \gls{map} symbol detector for multichannel transmission in the presence of arbitrarily correlated phase noise.
Moreover, the algorithm is designed to be used after adaptive equalization has taken place, i.e., the \gls{cpe} performed by the algorithm is not embedded in the loop of, e.g., a \gls{lms} equalizer. Finally, it can perform \gls{cpe} on any number of channels. Hence, although it is designed for \gls{jccpe}, it can also be used on a channel-by-channel basis to perform \gls{pccpe}.
In this section, the main derivation results are presented. For more details on the derivations, refer to \cite{alfredsson:tcom18_arxiv}, where an analogous bit-detection algorithm for coded transmission is derived using the same techniques.

Let $\bm r$, $\bm s$, and $\bm\theta$ contain all the received samples, transmitted symbols, and phase-noise samples, respectively. The \gls{map} symbol-detection strategy is optimal in the sense that it minimizes the symbol error rate \cite[Ch.~5.1]{proakis:dc}. It is performed on a symbol-by-symbol basis according to
\begin{equation}
	\hat s_{i,k} = \argmax_{x\in\modset}P(s_{i,k}|\bm r).
	\label{eq:map}
\end{equation}
The \textit{a posteriori} symbol probability in \eqref{eq:map} is hard to compute exactly for the considered system model. However, it can be expressed as the marginalization of the joint \textit{a posteriori} distribution of $\bm s$ and $\bm\theta$, i.e.,
\begin{equation}
	P(s_{i,k}|\bm r)=\int_{\mathbb R^{D\times N}}\sum_{\bm s\in\mathcal S_{i,k}}p(\bm s,\bm\theta|\bm r)\mathrm d\bm\theta,\label{eq:marg}
\end{equation}
where $\mathcal S_{i,k}=\{\bm s'\in\modset^{D\times N}:s_{i,k}'=s_{i,k}\}$. Moreover, $p(\bm s,\bm\theta|\bm r)$ can be factorized as
\begin{equation}
	p(\bm s,\bm\theta|\bm r)\propto \prod_{k=2}^N p(\bm\theta_k|\bm\theta_{k-1})\prod_{i,k} p(r_{i,k}|s_{i,k},\theta_{i,k}),\label{eq:margFact}
\end{equation}
where $\propto$ denotes proportionality with respect to $\bm r$.

Applying the sum--product algorithm on a factor graph associated with the right-hand side of \eqref{eq:margFact} yields messages $P_{i,k}(s_{i,k})$ and $p_k(\bm\theta_k)$ that approximate the \textit{a posteriori} probability of $s_{i,k}$ and probability density function of $\bm\theta_k$, resp., for all $i,k$. The factor graph does not contain any cycles, and hence, it does not yield an iterative algorithm. However, to improve performance, the messages $P_{i,k}(s_{i,k})$ and $p_k(\bm\theta_k)$ are computed in an iterative fashion. In each iteration, $p_k(\bm\theta_k)$ is first updated for all $k$ through the use of \gls{eks} \cite[Ch.~9]{sarkka:bayesian} and soft symbols, i.e., the first and second moments of $s_{i,k}$ with respect to $P_{i,k}(s_{i,k})$ from a previous iteration, for all $i,k$,. Then, $P_{i,k}(s_{i,k})$ is updated using the current estimates of $p_k(\bm\theta_k)$ for all $(i,k)\in\mathcal D$. This iterative process reduces the resulting \gls{ber} after symbol-to-bit mapping until the iterations converge, after which the algorithm performance cannot be improved further. The convergence speed depends on various system parameters such as the modulation format, pilot rate, and laser linewidth. In this paper, a fixed number of iterations is used as a criterion for stopping the iterations.

\begin{figure}[!t]
	\centering
	\includegraphics{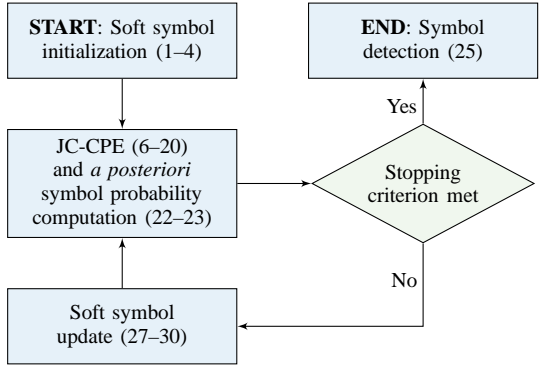}
	\caption{Flowchart illustrating the high-level structure of \gls{fgk}. The numbers in the blocks correspond to line numbers in Algorithm~\ref{alg:ijpt}.}
	\label{fig:flowchart}
\end{figure}

This \gls{cpe} algorithm will be referred to as \gls{fgk} hereafter. It is described in a high-level manner in Fig.~\ref{fig:flowchart} and detailed in the form of a pseudocode in Algorithm~\ref{alg:ijpt}. The first and second moments of $s_{i,k}$, denoted as $\bar s_{i,k}$ and $\bar\sigma_{i,k}^2$, are initialized in lines 1--4 such that $\bar s_{i,k}=\rho_{i,k}$ and $\bar\sigma_{i,k}^2=\sigma_i^2$ for all $(i,k)\in\mathcal P$, whereas $\bar s_{i,k}=0$ and $\bar\sigma_{i,k}^2=\sigma_i^2+E_\mathrm{s}/2$ for all $(i,k)\in\mathcal D$. The \gls{eks} equations, used to estimate $p_k(\bm\theta_k)$ for all $k$, are then listed in lines 6--20, where $\bm\theta_{k}^\mathrm{s}$ and $\mathbf M_k^\mathrm{s}$ represent the estimated mean and covariance of $p_k(\bm\theta_k)$ at time $k$, resp. Finally, the logarithm of $P_{i,k}(s_{i,k})$ is computed\footnote{The reason for computing the logarithm of $P_{i,k}(s_{i,k})$ is to ensure numerical stability.} for all $(i,k)\in\mathcal D$ in lines 22--23, where $\theta_{i,k}^\mathrm{s}$ is the $i$th component of $\bm\theta_k^\mathrm{s}$ and $M_{i,k}^\mathrm{s}$ is the $i$th element on the diagonal line of $\mathbf M_k^\mathrm{s}$. If the stopping criterion has been met, symbol detection is performed in line 25. Otherwise, the soft symbols are updated in lines 27--30 and another iteration is run.

\begin{algorithm}
	\caption{\gls{fgk}}
	\label{alg:ijpt}
	\begin{algorithmic}[1]
		\Require $\bm r$, $D$, $N$, $\mathcal P$, $\mathcal D$, $\mathbf Q$, $\bm\sigma^2$, $\modset$
		\Ensure $\hat s_{i,k},\forall~(i,k)\in\mathcal D$
			\State $\bar s_{i,k}=\rho_{i,k},\forall~(i,k)\in\mathcal P$
			\State $\bar s_{i,k}=0,\forall~(i,k)\in\mathcal D$
			\State $\varNtilde{i}=\varN{i},\forall~(i,k)\in\mathcal P$
			\State $\varNtilde{i}=\varN{i}+\Es/2,\forall~(i,k)\in\mathcal D$
		\ForAll{iterations}
			\State $\bm{\hat\theta}^{\mathrm f}_{1}=\left[\angle(r_{1,1}\bar s_{1,1}^*),\dots,\angle(r_{D,1}\bar s_{D,1}^*)\right]^T$
			\State $\mathbf M^{\mathrm f}_{1}=\diag(\varNtilde{1}/\Es,\dots,\varNtilde{D}/\Es)$
			\For{$k=2,\dots,N$}
				\State $h_{i,k}={\Im\{r_{i,k}\bar s_{i,k}^*e^{-j\hat\theta^{\mathrm f}_{i,k-1}}\}}/{\varNtilde{i}},\forall~i=1,\dots,D$
				\State $\mathbf V_k=\diag({\left|\bar s_{1,k}\right|^2}/{\varNtilde{1}},\dots,{\left|\bar s_{D,k}\right|^2}/{\varNtilde{D}})$
				\State $\mathbf M^{\mathrm f}_{k|k-1}=\mathbf M^{\mathrm f}_{k-1}+\mathbf Q$
				\State $\mathbf M^{\mathrm f}_{k}=(\mathbf I_D+\mathbf M^{\mathrm f}_{k|k-1}\mathbf V_k)^{-1}\mathbf M^{\mathrm f}_{k|k-1}$
				\State $\bm{\hat\theta}^{\mathrm f}_{k}=\bm{\hat\theta}^{\mathrm f}_{k-1}+\mathbf M^{\mathrm f}_{k}{\bm h}_k$
			\EndFor
			\State $\bm{\hat\theta}^{\mathrm s}_N=\bm{\hat\theta}^{\mathrm f}_{N}$ and $\mathbf M^{\mathrm s}_N=\mathbf M^{\mathrm f}_{N}$
			\For{$k=N-1,N-2,\dots,1$}
				\State $\mathbf A_k=\mathbf M^{\mathrm f}_{k}(\mathbf M^{\mathrm f}_{k+1|k})^{-1}$
				\State $\bm{\hat\theta}^{\mathrm s}_k=\bm{\hat\theta}^{\mathrm f}_{k}+\mathbf A_k(\bm{\hat\theta}^{\mathrm s}_{k+1}-\bm{\hat\theta}^{\mathrm f}_{k})$
				\State $\mathbf M^{\mathrm s}_k=\mathbf M^{\mathrm f}_{k}+\mathbf A_k(\mathbf M^{\mathrm s}_{k+1}-\mathbf M^{\mathrm f}_{k+1|k})\mathbf A_k^T$
			\EndFor
			\ForAll{$(i,k)\in\mathcal D$}
				\State $\xi_{i,k}(x)=\frac{e^{j\hat\theta_{i,k}^{\mathrm s}}}{M^{\mathrm s}_{i,k}}+\frac{r_{i,k}x^*}{\varN{i}}-\frac{r_{i,k}\bar s_{i,k}^*}{\varNtilde{i}},\forall~x\in\modset$
				\State $f_{i,k}(x)=|\xi_{i,k}(x)|-\frac{\left|x\right|^2}{2\varN{i}}-\frac12\ln|\xi_{i,k}(x)|,\forall~x\in\modset$
			\If{stopping criterion met}
				\State $\hat s_{i,k}=\argmax_{x\in\modset}f_{i,k}(x)$
			\Else
				\State $f_{i,k}^\mathrm{max}=\max_{x\in\modset} f_{i,k}(x)$
				\State $P_{i,k}(x)=\frac{\exp(f_{i,k}(x)-f_{i,k}^\mathrm{max})}{\sum_{x'\in\modset}\exp(f_{i,k}(x')-f_{i,k}^\mathrm{max})},\forall~x\in\modset$
				\State $\bar s_{i,k}=\sum_{x\in\modset}x P_{i,k}(x)$
				\State $\bar\sigma_{i,k}^2=\sigma_i^2+\frac12\sum_{x\in\modset}|x-\bar s_{i,k}|^2 P_{i,k}(x)$
			\EndIf
			\EndFor
		\EndFor
	\end{algorithmic}
\end{algorithm}

\section{Simulation Results}
\label{sec:simresults}

In this section, \gls{fgk} and \gls{bps} are first compared for \gls{pccpe} in terms of phase-noise tolerance in order to put the performance of \gls{fgk} into perspective.
Thereafter, \gls{pccpe} and \gls{jccpe}, which are both implemented using \gls{fgk}, are compared in terms of the resulting power efficiency, information rate, and laser-linewidth requirements of the system. To that end, uncoded transmission of Gray-mapped \gls{dpmqam}, for $M=16,64,256,1024$, is carried out using Monte Carlo simulations for different numbers of cores, pilot \glspl{oh}, laser linewidths, and degrees of spatial correlation in the phase noise. \gls{fgk} is run for 2 iterations in all cases unless otherwise stated.

\begin{figure}[!t]
	\centering
	\includegraphics{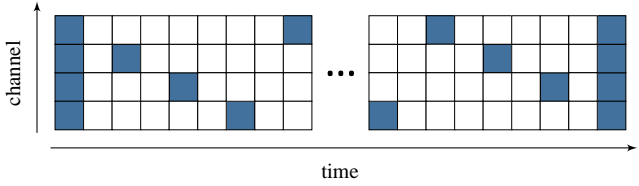}
	\caption{The pilot symbol arrangement where the pilot symbols, spaced at every other time index, are placed on a ``wrapped diagonal" line \cite{alfredsson2017ecoc}. Dark and white squares indicate pilot and data symbols, resp.}
	\label{fig:pd}
\end{figure}

The length of the transmitted block in each channel is $N=10\,000$ data and pilot symbols. The pilot symbols all take on the same point, i.e., $\rho_{i,k}=\sqrt{\Es}~\forall~(i,k)\in\mathcal P$. For \gls{pccpe}, they are distributed identically in each channel, i.e., uniformly throughout the transmitted block. For \gls{jccpe}, we showed in \cite{alfredsson2017ecoc} that a particular pilot symbol arrangement, where the pilot symbols are placed on a diagonal line that wraps around the top and bottom rows of the matrix, performs well in general for a wide range of phase-noise correlation degrees. The average pilot \gls{oh} across the channels, $\OHpilot$, can be adjusted by changing the temporal distance between the pilot symbols. Finally, all channels have a pilot symbol at the beginning and end of the transmitted block. This arrangement is illustrated in Fig.~\ref{fig:pd} for 4 spatial channels and the pilot symbols placed at every other time index.

\begin{figure}[!t]
	\centering
	\includegraphics{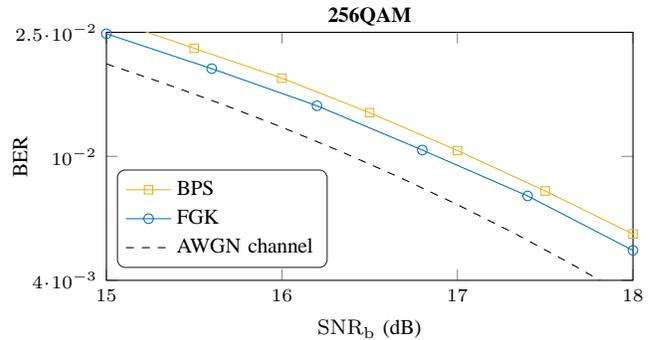}
	\caption{\gls{ber} versus $\snrinline$ for transmission of 256\gls{qam} in the presence of phase noise, comparing \gls{fgk} and \gls{bps} for \gls{pccpe}. As a reference, the theoretical \gls{ber} of transmission over the \gls{awgn} channel is included.}
	\label{fig:simBPSvsFGEKS}
\end{figure}

\begin{figure*}[!t]
	\centering
	\includegraphics{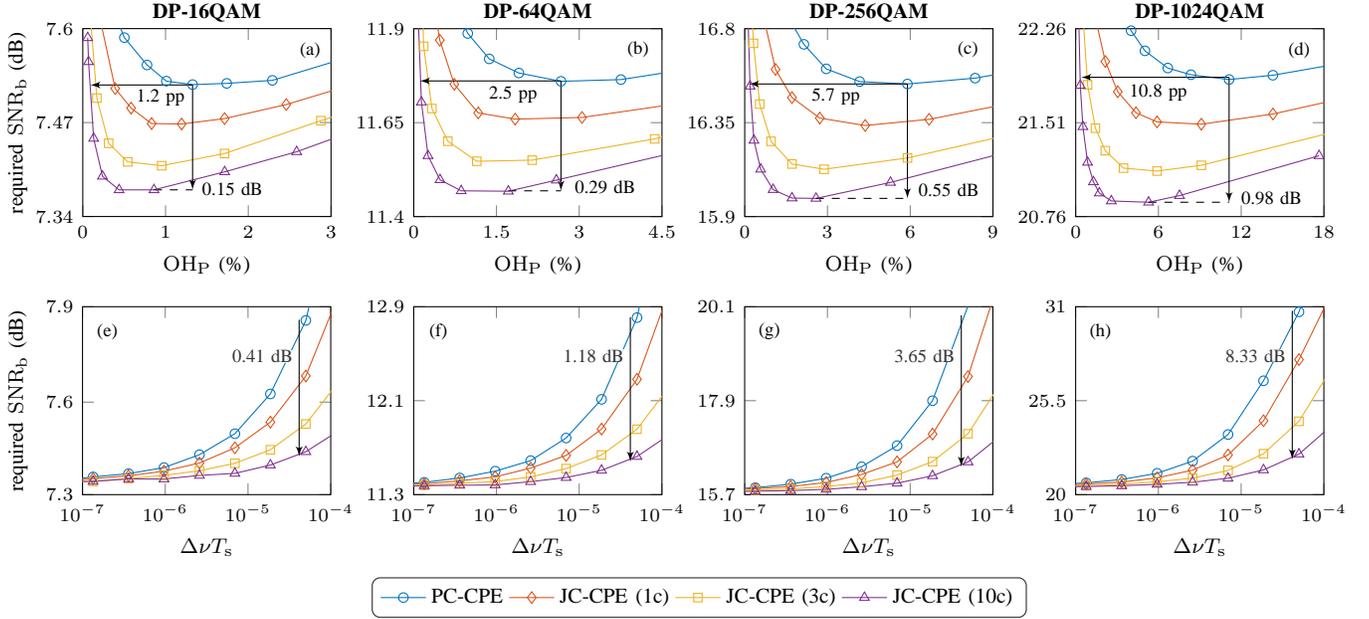}
	\caption{The required $\snrinline$ to attain the target pre-\gls{fec} \gls{ber} threshold of $1.44\cdot10^{-2}$ versus $\OHpilot$ (top) and $\Delta\nu T_\mathrm{s}$ (bottom), comparing \gls{pccpe} and \gls{jccpe} for different modulation formats and numbers of cores.}
	\label{fig:simPE}
\end{figure*}

Performance is assessed by estimating \gls{ber} or \gls{air} for different $\OHpilot$, laser linewidth and symbol duration products, $\Delta\nu T_\mathrm{s}$, and \glspl{snr} per bit \cite[Ch.~1]{heegard_turbocoding}, $\snrinline$, defined as
\begin{equation}
	\snr\define\frac{E_s(1+\OHpilot)}{2\varN{}\log_2|\modset|},\label{eq:ebn0}
\end{equation}
where $\varN{}$ is the complex \gls{awgn} variance per real dimension, which is identical for all channels in this section. From \eqref{eq:ebn0}, it can be seen that the rate reduction due to the use of pilots is penalized through an increased AWGN variance, since for a fixed $\Es$, $|\modset|$, and $\snr$, $\varN{}$ increases with $\OHpilot$. For each \gls{ber} estimate, bit errors are accumulated through repeated transmission until the total number of bit errors reaches at least $10^4$. The data bits, phase noise, and \gls{awgn} are realized according to the system model in \eqref{eq:sysmodel}, independently for each transmission.

\subsection{Comparison Between \gls{fgk} and \gls{bps} for \gls{pccpe}}
\label{sec:bps_comp}

To ensure high performance for both algorithms, \gls{bps} uses 128 test phases and the filter length is optimized for each tested \gls{snr} to minimize the resulting \gls{ber}, whereas \gls{fgk} is run for 20 iterations to allow for proper convergence of the \gls{cpe}. Moreover, \gls{bps} is provided with perfect knowledge of the initial value of the phase noise. Fig.~\ref{fig:simBPSvsFGEKS} shows \gls{ber} as a function of $\snrinline$ for transmission of 256\gls{qam} at 20 GBd, with $\Delta\nu=200$ kHz and $\OHpilot=1\%$ for \gls{fgk}. The theoretical \gls{ber} performance for uncoded, pilot-free transmission of Gray-mapped 256\gls{qam} over the \gls{awgn} channel \cite{1021039} is included as a reference. \gls{fgk} outperforms \gls{bps} across all tested \glspl{snr}, which is consistent with the coded-transmission results presented in \cite{alfredsson:tcom18_arxiv}.

\subsection{Power Efficiency}

To assess the gains in power efficiency that \gls{jccpe} enables, the required $\snrinline$ to achieve a target pre-\gls{fec} \gls{ber} threshold of $1.44\cdot10^{-2}$, corresponding to a staircase code with a \gls{fec} \gls{oh} of $\OHFEC=20\%$ \cite{Zhang:14}, is estimated for different $\OHpilot$, $\Delta\nu T_\mathrm{s}$, and degrees of spatial correlation in the phase noise. Since the system model in (1) does not account for nonlinearities, the results in this subsection are mostly relevant to short-haul transmission where the system operates in the linear regime.

Figs.~\ref{fig:simPE} (a)--(d) show the required $\snrinline$ as a function of $\OHpilot$ at 20 GBd for $\Delta\nu=200$ kHz, $\varCoreDrift=\varLPN/10^3$, and $\varPolDrift=\varLPN/10^6$. As can be seen, there exists an optimal $\OHpilot$ for all modulation formats, denoted with $\OHpilot^*$, that requires the minimum $\snrinline$ to attain the pre-\gls{fec} \gls{ber} threshold. The phase-noise tolerance of \gls{jccpe} increases with the number of cores, which leads to a lower minimum required $\snrinline$ compared to \gls{pccpe}. This reduction, marked with a vertical arrow in each plot, is up to 0.98 dB for transmission of \gls{dp}-1024\gls{qam} through 10 cores. However, note that $\OHpilot^*$ is higher for \gls{pccpe} than \gls{jccpe}. Moreover, the difference in required $\snrinline$ highly depends on $\OHpilot$. As an example, a difference of 0.15 dB, 0.41 dB, 1.12 dB, and 3.38 dB is observed at $\OHpilot=1\%$ for 10-core transmission of \gls{dp}-16\gls{qam}, \gls{dp}-64\gls{qam}, \gls{dp}-256\gls{qam}, \gls{dp}-1024\gls{qam}, resp.

Figs.~\ref{fig:simPE} (e)--(h) show the required $\snrinline$ as a function of $\Delta\nu T_\mathrm{s}$ for $\OHpilot=1\%$, $\varCoreDrift=\varLPN/10^3$, and $\varPolDrift=\varLPN/10^6$. As before, the phase-noise tolerance increases for \gls{jccpe} with the number of cores, and thus, the required $\snrinline$ to attain the pre-\gls{fec} \gls{ber} threshold is less for \gls{jccpe} than \gls{pccpe}. Furthermore, the difference in the required $\snrinline$ between \gls{pccpe} and \gls{jccpe} grows with increasing $\Delta\nu T_\mathrm{s}$, i.e., with increasing laser linewidth and/or decreasing symbol rate. The difference for $\Delta\nu T_\mathrm{s}=5\cdot10^{-5}$, again marked with a vertical arrow in the plots, is up to 8.33 dB in the case of \gls{dp}-1024\gls{qam} transmission through 10 cores. The value $\Delta\nu T_\mathrm{s}=5\cdot10^{-5}$ corresponds to, e.g., $\Delta\nu=1$ MHz at 20 GBd or $\Delta\nu=100$ kHz at 2 GBd.

\begin{figure}[!t]
	\centering
	\includegraphics{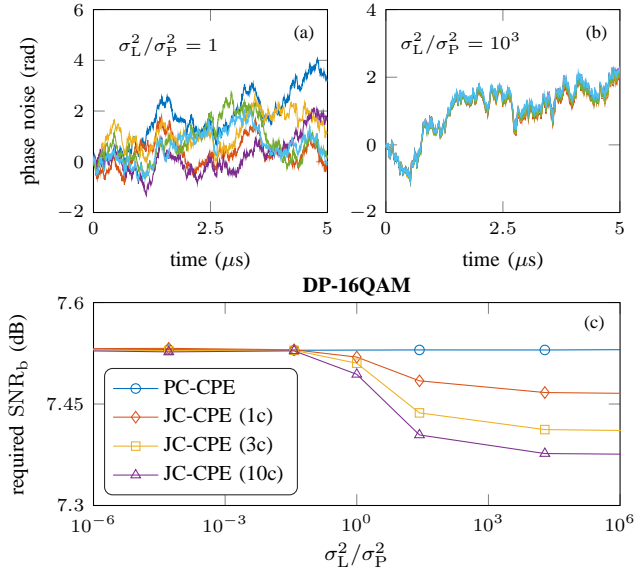}
	\caption{(a)--(b) Phase-noise realizations for transmission through 6 spatial channels with two different correlation degrees. (c) The required $\snrinline$ to attain the target pre-\gls{fec} \gls{ber} threshold of $1.44\cdot10^{-2}$ versus $\varLPN/\varPolDrift$, comparing \gls{pccpe} and \gls{jccpe} for different numbers of cores.}
	\label{fig:simPEcorr}
\end{figure}

For illustration purposes, Figs.~\ref{fig:simPEcorr} (a) and (b) show phase-noise realizations in 6 channels for two different degrees of phase-noise correlation, namely $\varLPN/\varPolDrift=1$ and $\varLPN/\varPolDrift=10^3$. Moreover, Fig.~\ref{fig:simPEcorr} (c) shows the required $\snrinline$ as a function of phase-noise correlation for transmission of \gls{dp}-16\gls{qam} with $\OHpilot=1\%$. The correlation is quantified by fixing $\varLPN+\varPolDrift=6.3\cdot10^{-5}$ (corresponding to $\Delta\nu=200$ kHz at 20 GBd) and $\varCoreDrift=0$, while varying the ratio $\varLPN/\varPolDrift$. As the ratio tends to 0, the phase noise becomes independent between the spatial channels, and \gls{jccpe} has no performance gain over \gls{pccpe}. However, as the ratio tends to infinity, the phase noise becomes identical in all channels. For $\varLPN/\varPolDrift>10^{-1}$, \gls{jccpe} outperforms \gls{pccpe}, and the performance gains grow with increasing correlation. At approximately $\varLPN/\varPolDrift=10^4$, \gls{jccpe} reaches a point where a greater correlation yields marginal gains. Identical results are found for \gls{dp}-64\gls{qam}, \gls{dp}-256\gls{qam}, and \gls{dp}-1024\gls{qam}.

\begin{figure*}[!t]
	\centering
	\includegraphics{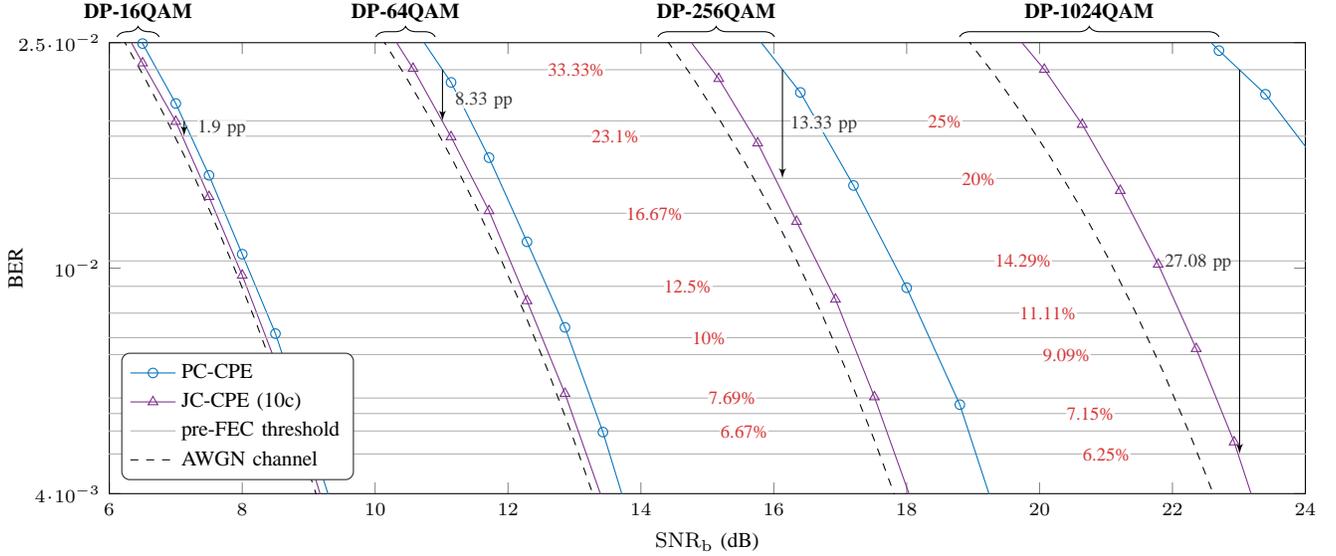}
	\caption{\gls{ber} versus $\snrinline$ for transmission of different modulation formats through 10 cores, comparing \gls{pccpe} and \gls{jccpe}. The pre-\gls{fec} \gls{ber} thresholds and the corresponding $\OHFEC$ of the considered staircase codes are marked in the plot with gray horizontal lines and red text, resp. As a reference, the theoretical \gls{ber} of transmission in the absence of phase noise is included. (pp: percentage point)}
	\label{fig:simSE}
\end{figure*}

\subsection{Information Rate}

The information rate of the system can be increased by, e.g., lowering $\OHFEC$ and/or $\OHpilot$. In Figs.~\ref{fig:simPE} (a)--(d), it can be seen that the value of $\OHpilot$ corresponding to a particular required $\snrinline$ decreases for \gls{jccpe} with an increasing number of cores. As an example, the difference in $\OHpilot$ between \gls{pccpe} and \gls{jccpe} to achieve the minimum required $\snrinline$ of \gls{pccpe}, marked with a horizontal arrow in each plot, is up to 10.8 percentage points for transmission of \gls{dp}-1024\gls{qam} through 10 cores. Given the assumption of $\OHFEC=20\%$, this corresponds to a 12.1\% rate increase (from 7.41 b/symbol to 8.31 b/symbol in each polarization).

Fig.~\ref{fig:simSE} shows \gls{ber} versus $\snrinline$, comparing \gls{pccpe} and \gls{jccpe} for transmission at 20 GBd through 10 cores with $\Delta\nu=200$ kHz, $\OHpilot=1\%$, $\varCoreDrift=\varLPN/10^3$, and $\varPolDrift=\varLPN/10^6$. To determine possible rate improvements that can be achieved by lowering $\OHFEC$ and performing \gls{jccpe}, several staircase codes from \cite{Zhang:14} are considered with their corresponding pre-\gls{fec} \gls{ber} thresholds and $\OHFEC$ values marked in the plot. As in Section \ref{sec:bps_comp}, the theoretical \gls{ber} performance for uncoded, pilot-free transmission of Gray-mapped \gls{qam} over the \gls{awgn} channel is included as a reference. As \gls{jccpe} yields lower \gls{ber} than \gls{pccpe}, it allows for the use of a \gls{fec} code with lower pre-\gls{fec} threshold and $\OHFEC$ than what could otherwise be used in the case of \gls{pccpe}. By considering $\snrinline$ values where the \gls{pccpe} performance crosses a pre-\gls{fec} \gls{ber} threshold, the greatest reduction in $\OHFEC$, marked with a vertical arrow for each modulation format at the corresponding $\snrinline$, is up to 27.08 percentage points in the case of \gls{dp}-1024\gls{qam} transmission. Since $\OHpilot=1\%$, this corresponds to a 25.5\% rate increase (from 7.50 b/symbol to 9.41 b/symbol in each polarization). Table~\ref{table:OHreduction} further details the maximum percentage-point reduction in $\OHFEC$ and the corresponding rate gain for different modulations formats and numbers of cores, again by looking at $\snrinline$ values where \gls{pccpe} crosses a pre-\gls{fec} \gls{ber} threshold. Overall, increasing the modulation format order and the number of cores yields a bigger rate improvement.

\begin{figure}[!t]
	\centering
	\includegraphics{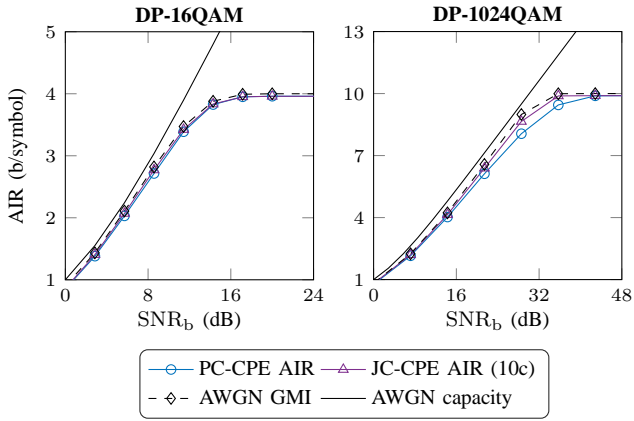}
	\caption{\gls{air} in bits per symbol per polarization versus $\snrinline$ for transmission of \gls{dp}-16\gls{qam} and \gls{dp}-1024\gls{qam} through 10 cores, comparing \gls{pccpe} and \gls{jccpe}. The channel capacity and GMI for the \gls{awgn} channel are included for reference.}
	\label{fig:simAIRvsSNR}
\end{figure}

In order to estimate possible rate improvements through \gls{jccpe} when
a soft-decision bit-wise \gls{fec} decoder is used, an appropriate performance metric to consider is a particular \gls{air}: the \gls{gmi} including rate loss due to pilot symbols. The \gls{gmi} is computed according to \cite[Eq.~(26)]{7138570} with the bit-wise log-likelihood ratios calculated in exact form. Fig.~\ref{fig:simAIRvsSNR} shows a comparison between \gls{pccpe} and \gls{jccpe} in terms of \gls{air} for different values of $\snrinline$. Transmission of \gls{dp}-16\gls{qam} and \gls{dp}-1024\gls{qam} at 20 GBd through 10 cores is considered, with $\Delta\nu=200$ kHz, $\OHpilot=1\%$, $\varCoreDrift=\varLPN/10^3$, and $\varPolDrift=\varLPN/10^6$. As a reference, the \gls{gmi} and Shannon capacity \cite{6773024} of the \gls{awgn} channel, are also shown. As Fig.~\ref{fig:simAIRvsSNR} shows, the performance gains due to joint processing are marginal for \gls{dp}-16\gls{qam} but significant for \gls{dp}-1024\gls{qam} with up to 0.59 b/symbol increase per polarization in \gls{air}.

\begin{table}[!t]
	\renewcommand{\arraystretch}{1.5}
	\caption{Maximum \gls{fec} \gls{oh} reduction and resulting SE increase for different modulation formats and numbers of cores (IR: information rate, pp: percentage point)}
	\label{table:OHreduction}
	\centering
	\begin{tabular}{r|cccccc}
		\hline\hline
		& \multicolumn{3}{c}{Max. \gls{oh} reduction (pp)} & \multicolumn{3}{c}{IR increase (\%)} \\
		& 1c & 3c & 10c & 1c & 3c & 10c \\
		\hline
		\gls{dp}-16\gls{qam} 		& 0		& 1.90 	& 1.90	& 0		& 1.54	& 1.54 \\
		\gls{dp}-64\gls{qam} 		& 1.90	& 3.33 	& 8.33	& 1.54 	& 2.85	& 6.66 \\
		\gls{dp}-256\gls{qam} 	& 4.29 	& 10.23	& 13.33	& 3.90 & 8.31 	& 11.11 \\
		\gls{dp}-1024\gls{qam} 	& 13.33	& 23.33	& 27.08	& 11.11	& 21.21	& 25.49 \\
		\hline\hline
	\end{tabular}
\end{table}

\subsection{Laser-Linewidth Requirements}

\begin{figure}[!t]
	\centering
	\includegraphics{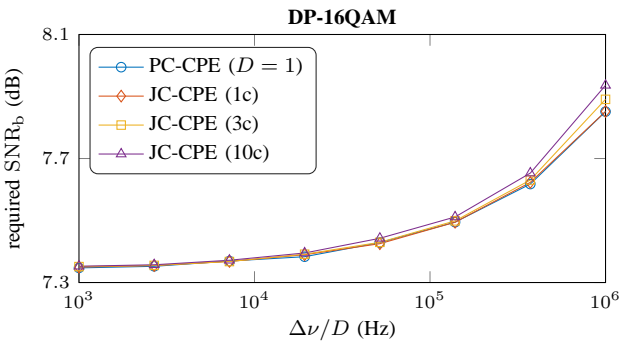}
	\caption{The required $\snrinline$ to attain the target pre-\gls{fec} \gls{ber} threshold of $1.44\cdot10^{-2}$ versus $\Delta\nu/D$, comparing \gls{pccpe} and \gls{jccpe} for different numbers of cores.}
	\label{fig:simPEnormLLW}
\end{figure}

Fig.~\ref{fig:simPEnormLLW} shows the required $\snrinline$ to attain a target pre-\gls{fec} \gls{ber} threshold of $1.44\cdot10^{-2}$ as a function of $\Delta\nu/D$, i.e., the laser linewidth normalized by the number of channels, for transmission of \gls{dp}-16\gls{qam} at 20 GBd with $\OHpilot=1\%$, $\varCoreDrift=\varLPN/10^3$, and $\varPolDrift=\varLPN/10^6$. All the curves essentially overlap, which shows that \gls{jccpe} for $D$ channels tolerates approximately $D$ times more laser linewidth compared to \gls{pccpe}. Thus, in the case of \gls{sdm} \gls{dp} transmission through $D/2$ cores that share lasers with linewidths $\Delta\nu$, \gls{jccpe} performs close to standard transmission through $D/2$ cores, where each core has an independent laser with linewidth $\Delta\nu/D$, provided that the phase noise has sufficient spatial correlation.

\begin{figure*}[!t]
	\centering
	\includegraphics{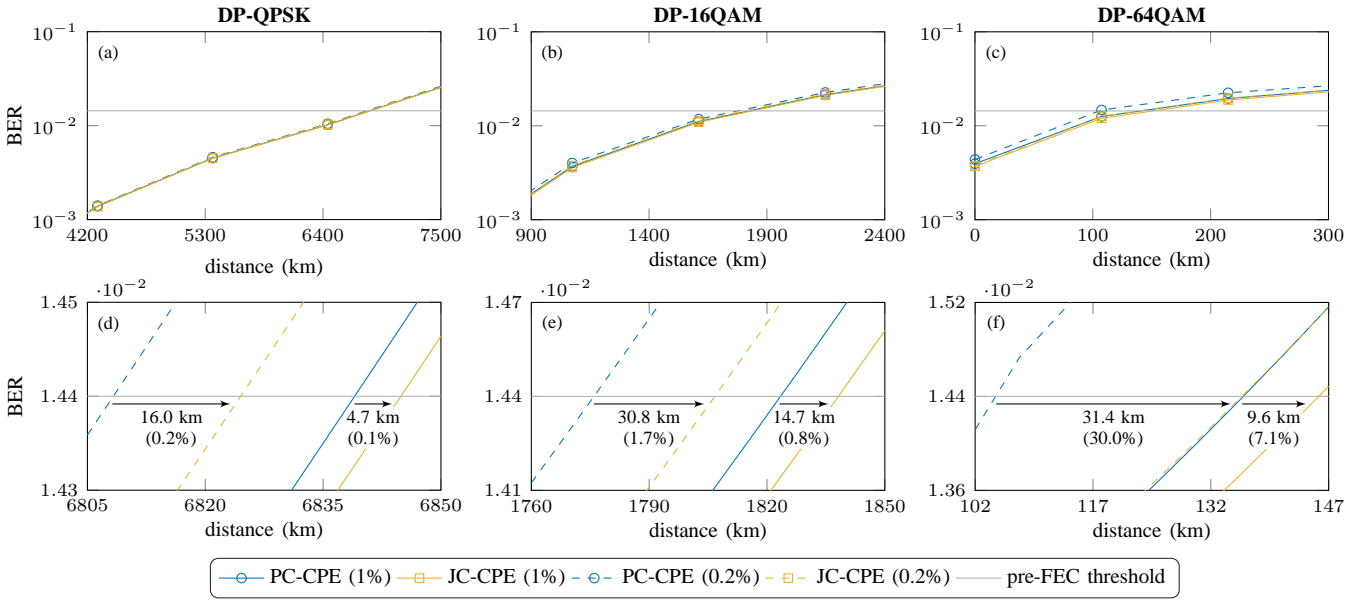}
	\caption{Experimental results showing \gls{ber} versus transmission distance, comparing \gls{jccpe} and \gls{pccpe} for transmission through 3 cores of a weakly-coupled single-mode \gls{mcf}, using different modulation formats and $\OHpilot$. The bottom plots show the same results zoomed in around the target pre-\gls{fec} \gls{ber} threshold of $1.44\cdot10^{-2}$.}
	\label{fig:expTD}
\end{figure*}

\section{Experimental Results}

In this section, \gls{pccpe} and \gls{jccpe} are experimentally compared in terms of \gls{ber} versus transmission distance, and the results are used to validate the system model in \eqref{eq:sysmodel}. The experimental setup consisted of 3 synchronized recirculating loops running through 3 adjacent cores of a 7-core, 53.7 km, weakly-coupled, single-mode, homogeneous \gls{mcf}. Transmission of \gls{dpqpsk}, \gls{dp}-16\gls{qam}, and \gls{dp}-64\gls{qam} was carried out at 20 GBd, with the corresponding transmitter output power set to $-4$ dBm, $-3.5$ dBm, and $0$ dBm, resp. An \gls{ecl} with 100 kHz linewidth operating at 1550 nm was used as a light source and shared for all cores at the transmitter, while a separate 100 kHz \gls{ecl} acting as an \gls{lo} was shared for all cores at the receiver. All detected signals were simultaneously digitized in a 12-channel oscilloscope operating at 80 GS/s for offline \gls{dsp}. For a more detailed description of the setup, refer to \cite{luis:ecoc17}.

All stages in the \gls{dsp} chain except the CPE were performed on a per-core basis. The first stages consisted of resampling to 2 samples per symbol, dispersion estimation and compensation, orthonormalization, timing recovery, and blind $2\times2$ equalization using the constant modulus algorithm. The orthonormalization was performed using the Gram--Schmidt algorithm, but alternatively, it could be achieved in the equalization stage through the use of a widely linear equalizer \cite{7486065}. At this point, frame synchronization was performed. A training stage followed where approximately $10^5$ symbols were used for \gls{lms} $2\times2$ equalization and estimation of the complex additive noise variance in each spatial channel. The rationale for performing per-core $2\times2$ equalization as opposed to $6\times6$ equalization was to save processing time, as the latter variant has a much higher complexity. Moreover, it was shown in \cite{luis:ecoc17} using the same setup as in this paper that per-core $2\times2$ equalization does not yield significantly worse performance than $6\times6$ equalization. After the training stage, an \gls{rde} with a slow convergence rate was used, initialized using the filter taps from the trained equalization. The \gls{rde} is immune to the signal phase and does not compensate for frequency offsets and phase noise. Therefore, an explicit carrier recovery stage was needed following the \gls{rde}. As \gls{fgk} is designed to be used after adaptive equalization, the choice the \gls{rde} is reasonable. The equalization was followed by blind \gls{cfo} compensation, down-conversion, matched filtering, and down-sampling to 1 sample per symbol.

Effective joint-core \gls{cfo} compensation was not possible due to the acousto-optic modulators used in the experimental setup, which introduced different frequency shifts to the signals running through the different cores. However, relative \glspl{cfo} between the cores were minimized as follows. Coarse \gls{cpe} was performed on a per-core basis using the Viterbi--Viterbi algorithm \cite{1056713}. The reason for using this algorithm was to save processing time and the fact that it sufficed for coarse \gls{cpe}. By subtracting the phase-noise estimates associated with a specific core from the estimates associated with the other two cores, the relative phases between the cores were obtained. Then, through linear least-squares fitting of the relative phases, the relative \glspl{cfo} were estimated and mitigated using the resulting fit. Moreover, a second orthonormalization step to correct for transmitter I/Q imbalances, which cannot be compensated before the carrier recovery stage due to the presence of \glspl{cfo} and phase noise, was carried out before symbol detection. Finally, each \gls{ber} estimate was computed by counting bit errors out of at least $10^7$ bits.

\gls{fgk} uses the covariance matrix $\mathbf Q$ in \eqref{eq:rwcov}, and hence, depends on $\varLPN$, $\varCoreDrift$, and $\varPolDrift$. These parameters were tweaked in order to minimize the resulting \gls{ber} estimates. For \gls{dpqpsk} and \gls{dp}-16\gls{qam}, different values of $\varLPN$ and $\varCoreDrift$ in the ranges $[50,300]$ kHz and $[\varLPN/10^4,\varLPN]$, resp., were found to minimize the \gls{ber}. As the phase noise was highly correlated in the two polarizations in each core, $\varPolDrift=10^6$ in all cases. Possible reasons for the wide ranges of optimal values for $\varLPN$ and $\varCoreDrift$ are different amounts of residual \glspl{cfo}, inaccurate estimates of the complex additive noise variance, and the presence of nonlinear phase noise. However, for \gls{dp}-64\gls{qam}, $\varLPN=40$ kHz, $\varCoreDrift=\varLPN/10^4$ and $\varPolDrift=\varLPN/10^6$ were overall the best out the tested values.

\begin{table}[!t]
	\renewcommand{\arraystretch}{1.5}
	\caption{Gains in total transmission reach for different modulation formats, pre-\gls{fec} \gls{ber} thresholds, and 1\% pilot \gls{oh}}
	\label{table:TDgain}
	\centering
	\begin{tabular}{c|ccc}
		\hline\hline
		\gls{ber} & \gls{dpqpsk} & \gls{dp}-16\gls{qam} & \gls{dp}-64\gls{qam} \\
		\hline
		$5.16\cdot10^{-3}$ & 5.5 km (0.1\%) & 13.4 km (1.1\%) & 3.9 km (21.0\%) \\
		$7.04\cdot10^{-3}$ & 5.4 km (0.1\%) & 11.8 km (0.8\%) & 3.4 km (8.5\%) \\
		$9.29\cdot10^{-3}$ & 5.4 km (0.1\%) & 10.3 km (0.7\%) & 6.5 km (10.0\%) \\
		$1.44\cdot10^{-2}$ & 4.7 km (0.1\%) & 14.7 km (0.8\%) & 9.6 km (7.1\%) \\
		$1.71\cdot10^{-2}$ & 4.7 km (0.1\%) & 14.0 km (0.7\%) & 11.9 km (6.8\%) \\
		$2.24\cdot10^{-2}$ & 4.7 km (0.1\%) & 17.9 km (0.8\%) & 15.5 km (5.5\%) \\
		\hline\hline
	\end{tabular}
\end{table}

Figs.~\ref{fig:expTD} (a)--(c) show the estimated \gls{ber} as a function of transmission distance for \gls{pccpe} and \gls{jccpe}, evaluated at $\OHpilot=1\%$ and $\OHpilot=0.2\%$, with a target pre-\gls{fec} \gls{ber} threshold of $1.44\cdot10^{-2}$. Moreover, Figs.~\ref{fig:expTD} (d)--(f) show the same results, zoomed in around the pre-\gls{fec} threshold. As can be seen, the performance difference between \gls{jccpe} and \gls{pccpe} grows with increasing modulation order and decreasing $\OHpilot$. The gains in transmission reach, marked with horizontal arrows in the plots, are determined in terms of absolute distance and percentage-wise gain. The largest percentage-wise gain of 30.0\% corresponds to the case of \gls{dp}-64\gls{qam} and $\OHpilot=0.2\%$. Furthermore, Table~\ref{table:TDgain} details the gains in transmission reach for $\OHpilot=1\%$ and different pre-\gls{fec} \gls{ber} thresholds, corresponding to selected staircase codes from \cite{Zhang:14}. For all the transmitted modulation formats, the percentage-wise gain in transmission reach is larger for lower pre-\gls{fec} \gls{ber} thresholds. Overall, the gains are negligible for \gls{dpqpsk} for \gls{dp}-16\gls{qam}, but in the case of \gls{dp}-64\gls{qam}, between 5.5\% and 21\% increased transmission reach is observed.

It is worth noting that these results depend on the signal quality following the DSP stages prior to the CPE. However, as mentioned above, all relative \glspl{cfo} were minimized before the \gls{cpe} was carried out, but skipping this step and letting \gls{fgk} track the residual \glspl{cfo} yields an insignificant performance penalty. For example, in the case of \gls{dp}-64\gls{qam} transmission using $\OHpilot=1\%$, the estimated \gls{ber} increases at most by 0.4\% (from $8.2\cdot10^{-3}$ to $8.23\cdot10^{-3}$).

\begin{figure}[!t]
	\centering
	\includegraphics{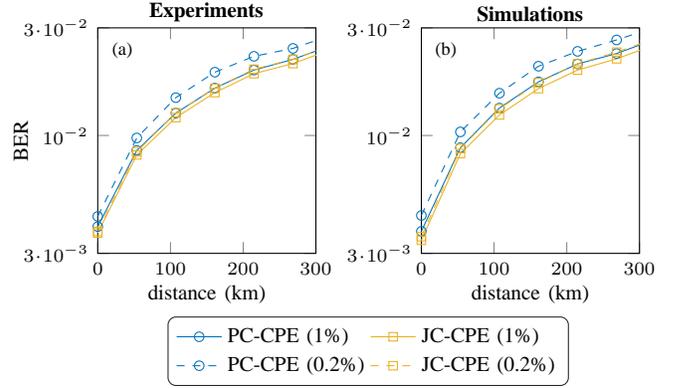}
	\caption{Comparison of (a) experimental and (b) Monte Carlo simulation results pertaining to \gls{ber} versus transmission distance for \gls{dp}-64\gls{qam} using $\OHpilot=1\%$ and $\OHpilot=0.2\%$.}
	\label{fig:expVSsim}
\end{figure}

The system model in \eqref{eq:sysmodel}, \eqref{eq:random_walk}, and \eqref{eq:rwcov}, which was used to develop \gls{fgk}, does not capture transmitter I/Q imbalances, and the random-walk assumptions for the different phase-noise components may not be accurate if, e.g., phase differences between the spatial channels are caused by residual \glspl{cfo} or nonlinearities. However, the transmission reach improvements suggest that the model is able to sufficiently describe the processed signal after all stages in the \gls{dsp} chain prior to the CPE. Moreover, Figs.~\ref{fig:expVSsim} (a) and (b) show the estimated \gls{ber} as a function of transmission distance for \gls{dp}-64\gls{qam} based on the experiments and simulations, resp. For the simulations, the \gls{awgn} variance was based on the estimated complex additive noise variance of the experimental data. Furthermore, $\varLPN=2\cdot10^{-6}$, which corresponds to $\Delta\nu=40$ kHz at 20 GBd, $\varCoreDrift=\varLPN/10^4$ and $\varPolDrift=\varLPN/10^6$. A strong agreement is observed between the simulations and the experimental results. Considering the same pre-\gls{fec} BER thresholds as in Table~\ref{table:TDgain}, the simulated transmission-reach gains in Fig.~\ref{fig:expVSsim} (b) range from 9.5\% to 24.3\% for $\OHpilot=1\%$, which is on the same order as the experimental gains.

\section{Conclusion}
\label{sec:concl}
The performance of \gls{jccpe} for \gls{sdm} transmission via \glspl{mcf} in the presence of correlated phase noise was studied. To that end, a phase-noise model was introduced that describes \gls{dp} transmission through an \gls{mcf}, where all cores share lasers on the transmitter and receiver sides, giving rise to a common \gls{lpn} in addition to core- and polarization-specific phase drifts, which decorrelate the phase noise across the spatial channels. It was further shown that this model can be regarded as a multidimensional random walk, and an algorithm developed for this model was used to compare the performance of \gls{pccpe} and \gls{jccpe} for \gls{mcf} transmission of various modulation formats through different numbers of cores, pilot \glspl{oh}, laser linewidths, and degrees of phase-noise correlation. Through Monte Carlo simulations of 20 GBd transmission with 200 kHz laser linewidth and 1\% pilot \gls{oh}, \gls{jccpe} led to a higher phase-noise tolerance. This could be exploited to improve power efficiency by up to 3.4 dB or increase information rate by up to 25.5\%. Furthermore, \gls{jccpe} can relax laser-linewidth requirements by up to a factor equal to the number of spatial channels. Finally, using data from \gls{mcf} transmission experiments, the system model was validated, and \gls{pccpe} and \gls{jccpe} were compared in terms of transmission reach. Up to 21\% increase in transmission distance was observed for 1\% pilot \gls{oh} through the use of \gls{jccpe}.


\begin{thebibliography}{10}
	\providecommand{\url}[1]{#1}
	\csname url@samestyle\endcsname
	\providecommand{\newblock}{\relax}
	\providecommand{\bibinfo}[2]{#2}
	\providecommand{\BIBentrySTDinterwordspacing}{\spaceskip=0pt\relax}
	\providecommand{\BIBentryALTinterwordstretchfactor}{4}
	\providecommand{\BIBentryALTinterwordspacing}{\spaceskip=\fontdimen2\font plus
		\BIBentryALTinterwordstretchfactor\fontdimen3\font minus
		\fontdimen4\font\relax}
	\providecommand{\BIBforeignlanguage}[2]{{%
			\expandafter\ifx\csname l@#1\endcsname\relax
			\typeout{** WARNING: IEEEtran.bst: No hyphenation pattern has been}%
			\typeout{** loaded for the language `#1'. Using the pattern for}%
			\typeout{** the default language instead.}%
			\else
			\language=\csname l@#1\endcsname
			\fi
			#2}}
	\providecommand{\BIBdecl}{\relax}
	\BIBdecl
	
	\bibitem{sdm:richardson13}
	D.~J. Richardson, J.~M. Fini, and L.~E. Nelson, ``Space-division multiplexing
	in optical fibres,'' \emph{Nat. Photon.}, vol.~7, no.~5, pp. 354--362, Apr.
	2013.
	
	\bibitem{6317137}
	M.~D. Feuer, L.~E. Nelson, X.~Zhou, S.~L. Woodward, R.~Isaac, B.~Zhu, T.~F.
	Taunay, M.~Fishteyn, J.~M. Fini, and M.~F. Yan, ``Joint digital signal
	processing receivers for spatial superchannels,'' \emph{{IEEE} Photon.
		Technol. Lett.}, vol.~24, no.~21, pp. 1957--1960, Nov. 2012.
	
	\bibitem{7121584}
	B.~J. Puttnam, R.~S. Lu\'{i}s, J.-M. {Delgado Mendinueta}, J.~Sakaguchi,
	W.~Klaus, Y.~Awaji, N.~Wada, A.~Kanno, and T.~Kawanishi, ``Long distance
	transmission in a multi-core fiber with self-homodyne detection,'' in
	\emph{Proc. Opt. Fiber Commun. Conf.}, Mar. 2015, p. Th1D.5.
	
	\bibitem{6533237}
	M.~D. Feuer, L.~E. Nelson, K.~Abedin, X.~Zhou, T.~F. Taunay, J.~F. Fini,
	B.~Zhu, R.~Isaac, R.~Harel, G.~Cohen, and D.~M. Marom, ``{ROADM} system for
	space division multiplexing with spatial superchannels,'' in \emph{Proc. Opt.
		Fiber Commun. Conf.}, Mar. 2013, p. PDP5B.8.
	
	\bibitem{Karlsson:17}
	M.~Karlsson and E.~Agrell, ``Multidimensional modulation and coding in optical
	transport,'' \emph{J. Lightw. Technol.}, vol.~35, no.~4, pp. 876--884, Feb.
	2017.
	
	\bibitem{7376999}
	P.~Poggiolini, A.~Nespola, Y.~Jiang, G.~Bosco, A.~Carena, L.~Bertignono, S.~M.
	Bilal, S.~Abrate, and F.~Forghieri, ``Analytical and experimental results on
	system maximum reach increase through symbol rate optimization,'' \emph{J.
		Lightw. Technol.}, vol.~34, no.~8, pp. 1872--1885, Apr. 2016.
	
	\bibitem{1056713}
	A.~J. Viterbi and A.~M. Viterbi, ``Nonlinear estimation of {PSK}-modulated
	carrier phase with application to burst digital transmission,'' \emph{{IEEE}
		Trans. Inf. Theory}, vol.~29, no.~4, pp. 543--551, Jul. 1983.
	
	\bibitem{4814758}
	T.~Pfau, S.~Hoffmann, and R.~No{\'e}, ``Hardware-efficient coherent digital
	receiver concept with feedforward carrier recovery for {$M$-QAM}
	constellations,'' \emph{J. Lightw. Technol.}, vol.~27, no.~8, pp. 989--999,
	Apr. 2009.
	
	\bibitem{4298982}
	E.~Ip and J.~M. Kahn, ``Feedforward carrier recovery for coherent optical
	communications,'' \emph{J. Lightw. Technol.}, vol.~25, no.~9, pp. 2675--2692,
	Sep. 2007.
	
	\bibitem{Cheng:13}
	H.~Cheng, Y.~Li, F.~Zhang, J.~Wu, J.~Lu, G.~Zhang, J.~Xu, and J.~Lin,
	``Pilot-symbols-aided cycle slip mitigation for {DP-16QAM} optical
	communication systems,'' \emph{Opt. Express}, vol.~21, no.~19, pp.
	22\,166--22\,172, Sep. 2013.
	
	\bibitem{6533070}
	M.~Morsy-Osman, Q.~Zhuge, M.~Chagnon, X.~Xu, and D.~V. Plant, ``Experimental
	demonstration of pilot-aided polarization recovery, frequency offset and
	phase noise mitigation,'' in \emph{Proc. Opt. Fiber Commun. Conf.}, Mar.
	2013, p. OTu3I.6.
	
	\bibitem{7341655}
	M.~Pajovic, D.~S. Millar, T.~Koike-Akino, R.~Maher, D.~Lavery, A.~Alvarado,
	M.~Paskov, K.~Kojima, K.~Parsons, B.~C. Thomsen, S.~J. Savory, and P.~Bayvel,
	``Experimental demonstration of multi-pilot aided carrier phase estimation
	for {DP-64QAM} and {DP-256QAM},'' in \emph{Proc. European Conf. Opt.
		Commun.}, Sep. 2015, p. Mo.4.3.3.
	
	\bibitem{Zhu:17}
	C.~Zhu and N.~Kaneda, ``Discrete cosine transform based pilot-aided phase noise
	estimation for high-order {QAM} coherent optical systems,'' in \emph{Proc.
		Opt. Fiber Commun. Conf.}, Mar. 2017, p. Th4C.1.
	
	\bibitem{Le:14}
	S.~T. Le, T.~Kanesan, M.~E. McCarthy, E.~Giacoumidis, I.~D. Phillips, M.~F.~C.
	Stephens, M.~Tan, N.~J. Doran, A.~D. Ellis, and S.~K. Turitsyn,
	``Experimental demonstration of data-dependent pilot-aided phase noise
	estimation for {CO-OFDM},'' in \emph{Proc. Opt. Fiber Commun. Conf.}, Mar.
	2014, p. Tu3G.4.
	
	\bibitem{6517220}
	R.~G.~H. van Uden, C.~M. Okonkwo, V.~A. J.~M. Sleiffer, M.~Kuschnerov,
	H.~de~Waardt, and A.~M.~J. Koonen, ``Single {DPLL} joint carrier phase
	compensation for few-mode fiber transmission,'' \emph{{IEEE} Photon. Technol.
		Lett.}, vol.~25, no.~14, pp. 1381--1384, Jul. 2013.
	
	\bibitem{LarsECOC:17}
	L.~Lundberg, M.~Mazur, A.~Lorences-Riesgo, M.~Karlsson, and P.~A. Andrekson,
	``Joint carrier recovery for {DSP} complexity reduction in frequency
	comb-based superchannel transceivers,'' in \emph{Proc. European Conf. Opt.
		Commun.}, Sep. 2017, p. Th.1.D.3.
	
	\bibitem{6290331}
	D.~V. Souto, B.-E. Olsson, C.~Larsson, and D.~A.~A. Mello, ``Joint-polarization
	and joint-subchannel carrier phase estimation for 16-{QAM} optical systems,''
	\emph{J. Lightw. Technol.}, vol.~30, no.~20, pp. 3185--3191, Oct. 2012.
	
	\bibitem{7183869}
	R.~S. Lu\'{i}s, B.~J. Puttnam, J.-M. {Delgado Mendinueta}, W.~Klaus, Y.~Awaji,
	and N.~Wada, ``Comparing inter-core skew fluctuations in multi-core and
	single-core fibers,'' in \emph{Proc. Conf. Lasers and Electro-Opt.}, May
	2015, p. SM2L.5.
	
	\bibitem{mazur_jlt18}
	M.~Mazur, A.~Lorences-Riesgo, J.~Schr\"{o}der, P.~A. Andrekson, and
	M.~Karlsson, ``High spectral efficiency {PM-128QAM} comb-based superchannel
	transmission enabled by a single shared optical pilot tone,'' \emph{J.
		Lightw. Technol.}, vol.~36, no.~6, pp. 1318--1325, Mar. 2018.
	
	\bibitem{agrell_cleo18}
	E.~Agrell, A.~F. Alfredsson, B.~J. Puttnam, R.~S. Lu\'is, G.~Rademacher, and
	M.~Karlsson, ``Modulation and detection for multicore superchannels with
	correlated phase noise,'' in \emph{Proc. Conf. Lasers and Electro-Opt.}, May
	2018, p. SM4C.3.
	
	\bibitem{aplp_puttnam18}
	B.~J. Puttnam, R.~S. Lu\'is, G.~Rademacher, A.~Alfredsson, W.~Klaus,
	J.~Sakaguchi, Y.~Awaji, E.~Agrell, and N.~Wada, ``Characteristics of
	homogeneous multi-core fibers for {SDM} transmission,'' \emph{APL Photon.
		J.}, vol.~4, no.~2, p. 022804, Feb. 2019.
	
	\bibitem{1400262}
	T.~C.~W. Schenk, X.-J. Tao, P.~F.~M. Smulders, and E.~R. Fledderus, ``Influence
	and suppression of phase noise in multi-antenna {OFDM},'' in \emph{Proc.
		Vehicular Technol. Conf.}, vol.~2, Sep. 2004, pp. 1443--1447.
	
	\bibitem{7080901}
	R.~Krishnan, G.~Colavolpe, A.~{Graell i Amat}, and T.~Eriksson, ``Algorithms
	for joint phase estimation and decoding for {MIMO} systems in the presence of
	phase noise and quasi-static fading channels,'' \emph{{IEEE} Trans. Signal
		Process.}, vol.~63, no.~13, pp. 3360--3375, Jul. 2015.
	
	\bibitem{alfredsson:tcom18_arxiv}
	\BIBentryALTinterwordspacing
	A.~F. Alfredsson, E.~Agrell, and H.~Wymeersch, ``Iterative decoding and
	phase-noise compensation for multichannel optical transmission,'' \emph{arXiv
		e-prints}, Dec. 2018. [Online]. Available: \url{arxiv.org/abs/1804.02263}
	\BIBentrySTDinterwordspacing
	
	\bibitem{alfredsson2017ofc}
	A.~F. Alfredsson, E.~Agrell, H.~Wymeersch, and M.~Karlsson, ``Phase-noise
	compensation for spatial-division multiplexed transmission,'' in \emph{Proc.
		Opt. Fiber Commun. Conf.}, Mar. 2017, p. Th4C.7.
	
	\bibitem{alfredsson2018sum}
	A.~F. Alfredsson, E.~Agrell, H.~Wymeersch, B.~J. Puttnam, and R.~S. Lu\'is,
	``Joint phase tracking for multicore transmission with correlated phase
	noise,'' in \emph{Proc. {IEEE} Summer Topicals Meeting Series}, Jul. 2018, p.
	MF1.2.
	
	\bibitem{4909145}
	I.~Fatadin, D.~Ives, and S.~J. Savory, ``Blind equalization and carrier phase
	recovery in a {16-QAM} optical coherent system,'' \emph{J. Lightw. Technol.},
	vol.~27, no.~15, pp. 3042--3049, Aug. 2009.
	
	\bibitem{proakis:dc}
	J.~G. Proakis, \emph{Digital communications}, 4th~ed.\hskip 1em plus 0.5em
	minus 0.4em\relax Boston, MA, USA: McGraw-Hill, 2000.
	
	\bibitem{sarkka:bayesian}
	S.~S\"{a}rkk\"{a}, \emph{Bayesian Filtering and Smoothing}, 1st~ed.\hskip 1em
	plus 0.5em minus 0.4em\relax Cambridge, UK: Cambridge University Press, 2013.
	
	\bibitem{alfredsson2017ecoc}
	A.~F. Alfredsson, E.~Agrell, H.~Wymeersch, and M.~Karlsson, ``Pilot
	distributions for phase tracking in space-division multiplexed systems,'' in
	\emph{Proc. European Conf. Opt. Commun.}, Sep. 2017, p. P1.SC3.48.
	
	\bibitem{heegard_turbocoding}
	C.~Heegard and S.~B. Wicker, \emph{Turbo Coding}, 1st~ed.\hskip 1em plus 0.5em
	minus 0.4em\relax Kluwer Academic Publishers, 1999.
	
	\bibitem{1021039}
	K.~Cho and D.~Yoon, ``On the general {BER} expression of one- and
	two-dimensional amplitude modulations,'' \emph{{IEEE} Trans. Commun.},
	vol.~50, no.~7, pp. 1074--1080, Jul. 2002.
	
	\bibitem{Zhang:14}
	L.~M. Zhang and F.~R. Kschischang, ``Staircase codes with 6\% to 33\%
	overhead,'' \emph{J. Lightw. Technol.}, vol.~32, no.~10, pp. 1999--2002, May
	2014.
	
	\bibitem{7138570}
	A.~Alvarado, E.~Agrell, D.~Lavery, R.~Maher, and P.~Bayvel, ``Replacing the
	soft-decision {FEC} limit paradigm in the design of optical communication
	systems,'' \emph{J. Lightw. Technol.}, vol.~33, no.~20, pp. 4338--4352, Oct.
	2015.
	
	\bibitem{6773024}
	C.~E. Shannon, ``A mathematical theory of communication,'' \emph{Bell Syst.
		Technical J.}, vol.~27, no.~3, pp. 379--423, Jul. 1948.
	
	\bibitem{luis:ecoc17}
	R.~S. Lu\'is, B.~J. Puttnam, G.~Rademacher, Y.~Awaji, and N.~Wada, ``On the use
	of high-order {MIMO} for long-distance homogeneous single-mode multicore
	fiber transmission,'' in \emph{Proc. Opt. Fiber Commun. Conf.}, Sep. 2017, p.
	Th.2.F.2.
	
	\bibitem{7486065}
	E.~P. da~Silva and D.~Zibar, ``Widely linear equalization for {IQ} imbalance
	and skew compensation in optical coherent receivers,'' \emph{J. Lightw.
		Technol.}, vol.~34, no.~15, pp. 3577--3586, Aug. 2016.
	
\end{thebibliography}


\end{document}